\documentstyle[epsf,twocolumn,prl,aps,floats]{revtex}
\newcommand{\figwidth}{3.375in} % Use for two column output
\begin{document}
\draft

% comment out the following line for single column output
 \twocolumn[\hsize\textwidth\columnwidth\hsize\csname @twocolumnfalse\endcsname

\title{Stripes and the
%two-dimensional
{\em t-J} model}
\author{ C. Stephen Hellberg\cite{steve} and E. Manousakis\cite{stratos} }
\address{ $^\dagger$Complex Systems Theory Branch,
Naval Research Laboratory, Washington, DC 20375}
\address{ $^\ddagger$Department of Physics and
%Center for Materials Research and Technology, \\
MARTECH,
Florida State University, Tallahassee, FL 32306-3016 }
\date{
%Submitted to Physical Review Letters
\today
}
\maketitle
\begin{abstract}
\noindent
We investigate the two-dimensional \mbox{\em t-J} model
at a hole doping of $x = \frac{1}{8}$ and $J/t=0.35$ with exact 
diagonalization.
The low-energy states are uniform (not striped).
We find numerous excited states with charge density wave structures,
which may be 
interpreted as striped phases.
%states of the pure {\em t-J} which does not take into account the long-range 
%part of the Coulomb interaction.
Some of these are consistent with 
neutron scattering data on the cuprates and nickelates.
\end{abstract}
\pacs{PACS numbers:  71.10.Fd, 74.20.Mn, 71.10.Pm}

% comment out the following line for single column output
 ]

In the search for an understanding of the 
cuprate
%copper-oxide
superconductors it is desirable to find a model which 
captures many of the essential aspects of the environment experienced by the 
electrons in these materials.
Because these materials are born out of 
antiferromagnetic insulators
%\cite{manousakis91}
by doping, it is somewhat urgent 
to decide if a simple model which begins from the strong electron 
correlation limit, such as the so-called {\em t-J} model, can explain some 
features which result from the electronic degrees of freedom in the
cuprates\cite{manousakis91}.
Even though the progress made in trying to solve the {\em t-J} model may be
characterized as slow, it gives some features 
which are present in these materials. For example, some important aspects of 
the calculated single-hole spectrum\cite{liu} are in agreement 
with the results 
of the photo-emission data\cite{wells}. In addition, the model  gives rise to 
a two-hole bound state\cite{boninsegni93} with the $d_{x^2-y^2}$ 
symmetry which is the believed symmetry of the superconducting 
state in these materials.

Emery and Kivelson\cite{emery90} suggested that the cuprates are near 
an electronic phase separation instability which is prevented by the 
long-range part of the Coulomb interaction.
In the phase-separated state, the holes cluster together,
leaving the rest of the system in an antiferromagnet state with no holes.
%On the other hand,
Phase separation in the {\em t-J} model
has been studied by a number of techniques which seem to be giving conflicting
conclusions\cite{hellberg97,exactdiag,putikka,kohno,sorella}.
For example, 
Hellberg and Manousakis\cite{hellberg97} 
using a stochastic 
projection method, an extension of the Green's function Monte Carlo (GFMC)
for lattice fermions,
find that the {\em t-J} model has a region of phase 
separation at all interaction strengths.  Other techniques fail to 
reach this conclusion.
Most of these studies use small size 
systems\cite{exactdiag},
high temperature series expansions\cite{putikka}, or
approximate methods\cite{kohno,sorella}.
% which effectively
%limits the length for non-zero correlations to half the order of the expansion
%and thus, they find a value of $ J/t \sim  1.2$ below which they find no 
%phase separation.
%However, 
%the energy scale for phase separation for small enough {\em J/t} is  
%smaller than the contribution to the energy from such finite-size effects.
In a very recent  calculation Calandra et al.\cite{sorella}  using the GFMC
approach  within the
fixed node approximation find that the phase boundary for phase separation is 
far from that determined
by the high temperature series expansions \cite{putikka}
and much closer to that obtained by Hellberg and Manousakis \cite{hellberg97}
except in the delicate region with small hole dopings and $J/t \lesssim 0.4$.
%except that for $J/t$ in the delicate region below $\sim 0.4$,
%they do not find phase separation.
%However,
By using 
a uniform Fermi-liquid type nodal structure one disregards the possibility of 
a non-uniform ground state in which one component of the mixture
(the antiferromagnetic 
phase) has no fermion degrees of freedom.
In addition, in the delicate region of
small $J/t$ and low doping, Shraiman and Siggia\cite{siggia} and 
Boninsegni and Manousakis\cite{boninsegni92}
showed spin-back-flow effects become very important resulting in the
interesting structure of the hole ``polaron''.
These effects are known to change the nodal structure
of the wave function in a crucial way in strongly correlated quantum fluids.
Therefore fixed-node GFMC may be inadequate in this region. 

%Hellberg and Manousakis\cite{hellberg97}, in agreement with 
%Emery and Kivelson\cite{emery93} concluded that
The phase separation in the {\em t-J} model cannot be realized
in the physical system due to the
%is the long-range part of the
Coulomb interaction\cite{emery90,hellberg97,emery93}.
Instead such a tendency for phase separation can be satisfied 
locally by forming stripes or other charge density wave (CDW)
structures without a large Coulomb cost.
%Additionally,
The coupling of electrons to
lattice distortions may also encourage the formation
of stripes.

Experimentally,
stripe modulations were first observed in a doped nickelate analogue
of the cuprates \cite{nickelates}.
La$_2$NiO$_4$ may be doped with holes by adding oxygen or by
substituting strontium for lanthanum.
The modulation seen with neutron scattering in the doped compounds is consistent
with the holes forming {\em diagonal} domain walls
separating antiferromagnetic regions of spins.
% with reduced hole density.
Stripes with a variety of widths and
%A variety of
hole densities along the stripes have been observed.
%A variety of diagonal stripes have been seen in these compounds
%with varying hole density along the stripes.

There is strong evidence for stripe modulations in the cuprates
as well\cite{cuprates}.
In La$_{1.6-x}$Nd$_{0.4}$Sr$_x$CuO$_4$, superconductivity
is suppressed at a filling of $x=\frac{1}{8}$, and neutron scattering
studies reveal {\em vertical} domain walls
of holes and spins.
In these stripes, a hole density of $\rho_h = 1/2$ per lattice spacing
is observed.
 
%More recently a particularly surprising possibility was suggested by 
Recently
White and Scalapino (WS) found static vertical stripe order
similar to that of the cuprates
in the two-dimensional {\em t-J} model
% at a filling of $x=\frac{1}{8}$
using a density matrix renormalization group
%(DMGR)
technique
%they find that the ground state of the {\em t-J} 
%model is a state characterized by striped order.
\cite{white98}.
These results are 
surprising due to the fact that the {\em t-J} model ignores
the long-range part of the Coulomb interaction and
couples to no lattice distortions.
One sees no 
physical reason for such a simplified model to have a ground state with
a periodic array of interfaces. 

In this paper we study the two-dimensional \mbox{\em t-J} model
at a hole doping of $x = \frac{1}{8}$ and at $J/t=0.35$
%(same parameter values used by WS) 
with exact diagonalization.
%At this point, the model is in the uniform phase,
%but is on the verge of phase separation.
The low-energy states are uniform (not striped).
We find a variety of CDW excited states
%We find a variety of relatively low-lying excited states
%which are CDW states
which may be interpreted as striped phases.
Some of these are in excellent agreement
with neutron scattering data on the cuprates and the nickelates.
However, without adding additional terms to the {\em t-J} model,
such as a long-range Coulomb interaction
or a coupling to the lattice, the striped
states are only realized as excited states.

%We argue that the findings of WS are artifact of the 
%choice of the boundary conditions and perhaps difficulties which 
%prevent the DMRG method from converging to the true ground state.
%In particular we argue that the stripes states which WS find are 
%excited states of a particular finite-size system defined by its 
%unit cell, the two necessary primitive vectors needed to tile the 
%two-dimensional space,
%the choice of boundary conditions as well as the amount of doping used. 
%In addition, we argue that the true ground state of the {\em t-J} 
%model for doping $1 \over 8$ and for $J/t=0.35$ (the values used  by WS), 
%is uniform in the thermodynamic limit.

%The Heisenberg model describes the experimental results of the undoped
%cuprate compounds well. 
%To describe the doped materials, it is desirable to
%extend the Heisenberg model to include mobile holes.
%The \mbox{\em t-J} model provides the
%simplest abstraction to describe the 
%environment experienced by the holes in the limit of strong
%on-site Coulomb repulsion. 
The \mbox{\em t-J} Hamiltonian
%, on a square lattice,
is written in the subspace with no doubly occupied sites as
\begin{equation}
   H
   = - t
      \sum_{\langle ij \rangle \sigma}
      ( c_{i\sigma}^{\dagger}c_{j\sigma}
		+ {\rm h.c.}
		)
 +
      J
\sum_{\langle ij \rangle}
      ( {\bf S}_{i} \! \cdot \! {\bf S}_{j} -
         \frac{n_i n_j}{4} ) .
\label{tj-ham}
\end{equation}
Here $\langle ij \rangle$ enumerates neighboring sites
on a square lattice,
$c_{i\sigma}^{\dagger}$ creates an electron of spin $\sigma$ on site
$i$, $n_i = \sum_\sigma c_{i\sigma}^{\dagger}c_{i\sigma}$,
and ${\bf S}_{i}$ is the spin-$\frac{1}{2}$ operator.
Throughout this paper, we take $t = 1$ and $J = 0.35$.

To achieve a doping of $x = \frac{1}{8}$,
all calculations were carried out on periodic 16-site clusters with two holes.
Periodic clusters may be characterized by their primitive translation
vectors {\bf a$_1$} and {\bf a$_2$}.
There are a large number of possible 16-site clusters on 
the two-dimensional square lattice.
Each cluster can only support striped phases that are
commensurate with the periodicity of that particular cluster\cite{prelovsek93}.
Clusters which have one particularly short translation vector
are quasi-one-dimensional
and behave like chains or ladders.
We only consider clusters in which both translation vectors have a
Manhattan length of at least $l_M = |a^x| + |a^y| \geq 4$.
There are seven such clusters, shown in Table \ref{table:clusters}.
These clusters represent all possible quasi-two-dimensional 16-site clusters.
Later in the paper, we examine six eigenstates in detail.
We label these states by letters, (a) through (f), shown
next to their corresponding clusters in Table \ref{table:clusters}.

\begin{table}[htb]
\caption{Translation vectors of the seven possible quasi-two-dimensional
16-site clusters.
The letters label eigenstates examined later in the paper.}
\label{table:clusters}
\begin{tabular}{cccc}
Number & {\bf a$_1$} & {\bf a$_2$} & States
\\ \hline 
1 & (0,4) & (4,0) & \\
2 & (0,4) & (4,1) & (e),(f) \\
3 & (0,4) & (4,2) & \\
4 & (3,2) & (2,-4) & \\
5 & (3,1) & (1,-5) & (d) \\
6 & (2,2) & (4,-4) & (a),(b)\\
7 & (2,2) & (3,-5) & (c)
\end{tabular}
\end{table}

We studied all eigenstates with energy per site
${\cal E} < -0.634$ for each of the seven periodic clusters.
For each cluster, we use all possible combinations of
phases $\theta = 0,\pi$ along each translation vector.
Several of these states are CDWs, which
are necessarily degenerate.
%In our case, the hole density on site {\bf r}$_i$ is given by
The density in a CDW state
is characterized by an amplitude $A$, a wave-vector {\bf k},
and an arbitrary phase $\phi$.
The phase $\phi$ depends on the particular linear combination of degenerate
states taken.
Thus the hole density on site ${\bf r}_i$ is given by
\begin{equation}
n_h({\bf r}_i) = \bar{n}_h + A \cos ( {\bf k} \cdot {\bf r}_i + \phi),
\label{density}
\end{equation}
where the average hole density is $\bar{n}_h = \frac{1}{8}$.
%The phase $\phi$ depends on the particular linear combination of degenerate
%states, and {\bf k} is the reciprocal lattice vector of the CDW.
%For our case of 14 electrons in a 16-site cluster one can  
%easily see the meaning of the relationship (\ref{density}).  There are only 
%two charge degrees of freedom in this case which are the position of each hole.
%Their center of mass momentum can only take one of the sixteen possible values
%while there is a distribution of the relative coordinate. Thus, 
%Eq. (\ref{density}) refers to the distribution of  the center of mass
%of a standing wave characterized by one of the 16 such momenta.
 
\begin{figure}
\epsfxsize=\figwidth\centerline{\epsffile{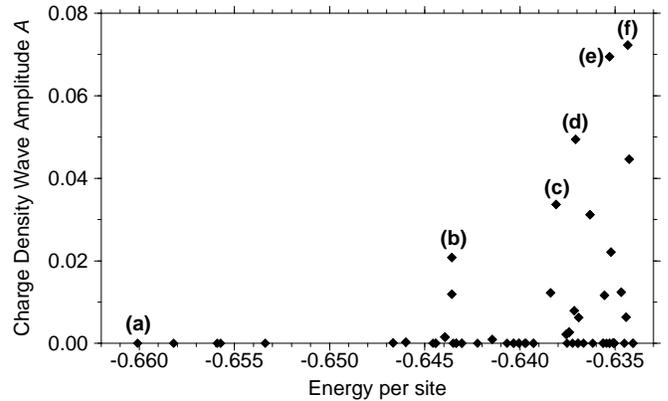}}
\vspace{.1in}
\caption{
Amplitude of the CDW in every state with energy per site ${\cal E} < -0.634$.
The six labeled states correspond to states examined in detail
later in the paper.
The low-energy states have no CDWs.
}
\label{fig:amplitude}
\end{figure}

The amplitude $A$ of the CDW in every low-energy state
are plotted in Fig.\ \ref{fig:amplitude} as a function of energy per site.
The lowest energy states are uniform and have CDW amplitude $A=0$.
For energies above ${\cal E} \gtrsim -0.645$
some CDW states are stabilized.
The maximum CDW amplitude of these states increases with increasing energy.
% This amplitude is bounded by the hole density $A \leq x = 0.125$.

\begin{figure}
\epsfxsize=\figwidth\centerline{\epsffile{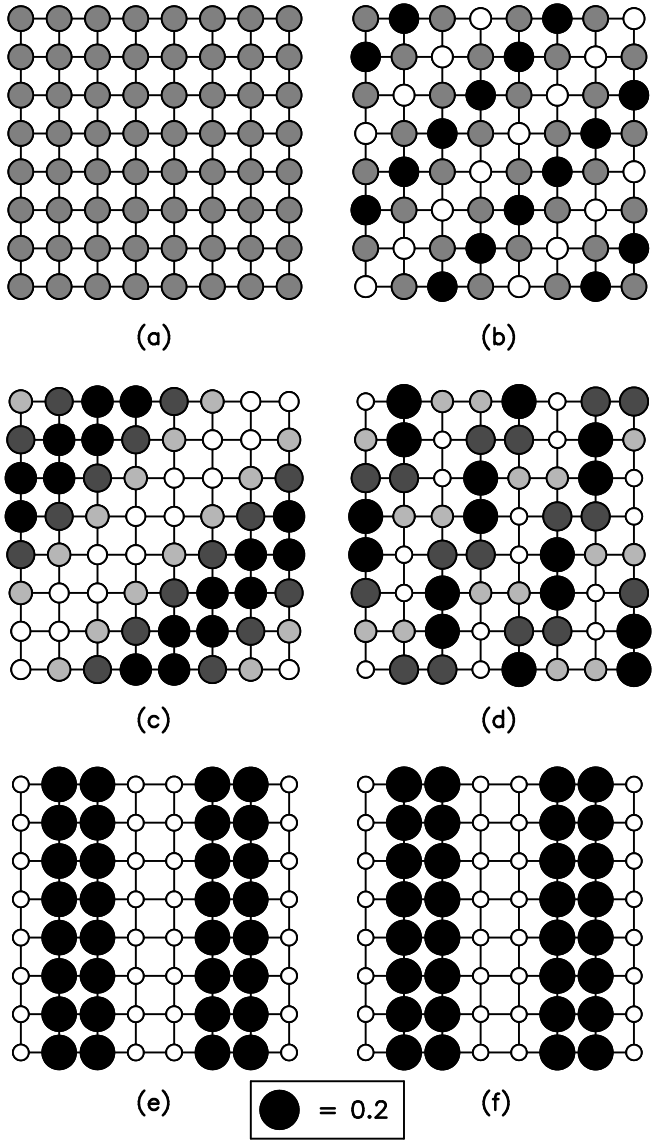}}
\vspace{.1in}
\caption{
Hole structures of the six eigenstates we chose to examine in detail.
The radius of each circle is proportional to the hole density
on the given site.
Additionally, the circles are shaded according to the relative
hole density in each state:
Black circles show the maximal hole density in that state
while white circles show the minimum.
State (a), which has the minimum energy, has a uniform hole density.
States (b) through (f) have increasing energy and show
increasing CDW amplitude.
In all states the average spin moment on each site is zero.
}
\label{fig:charge}
\end{figure}

%We labeled six states in Fig.\ \ref{fig:amplitude} that we examine in detail.
We examine in detail the six labeled states in Fig.\ \ref{fig:amplitude}.
State (a) is the lowest energy state and is uniform.
States (b) through (f) have increasing CDW amplitude
and increasing energy.
Each of these states has
the largest CDW amplitude of all states at or below its energy.

We examine the charge order of the six states in
Fig.\ \ref{fig:charge}.
The CDW states, (b) through (f), are degenerate, so taking a different linear
combination of the eigenstates will move the CDW.
In particular, in state (b) the maximum charge order occurs on a site,
while in states (c), (d), (e), and (f), the maximum charge order occurs
between two sites, or on a bond.  All of the states have site-centered 
and bond-centered CDWs with different linear combinations.

States (b) and (c) have diagonal stripes with two different
hole densities along the stripe.
State (b) has hole density
$\rho_b = 1/2$ per (1,1) step,
while state (c) has $\rho_c = 1$.
These states are similar to experimental results
on the nickelates \cite{nickelates} and to the mean-field calculations of
Zaanen and Littlewood\cite{zaanen94}.

States (e) and (f) exhibit vertical stripes with 1/2 hole
per (0,1) step consistent with 
experimental results on
%the cuprates
La$_{1.6-x}$Nd$_{0.4}$Sr$_x$CuO$_4$
\cite{cuprates} and with
the calculations of WS\cite{white98}.
States (e) and (f) have the largest CDW amplitudes that we found.

\begin{figure}[htb]
\epsfxsize=\figwidth\centerline{\epsffile{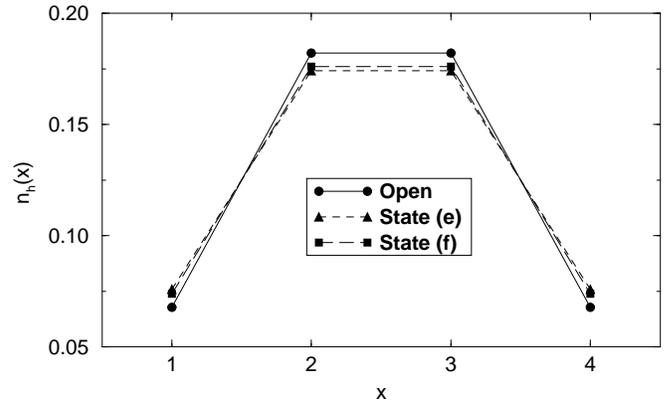}}
\vspace{.1in}
\caption{
Comparison of the hole density $n_h$ as a function of
$x$-coordinate for a $4 \times 4$ cluster with open boundary
conditions in the $x$-direction and states (e) and (f).
All three systems have exhibit a stripe in the $y$-direction.
The state with open boundary conditions is the non-degenerate ground state.
}
\label{fig:open}
\end{figure}

Interestingly, the CDWs of the vertical stripe states, (e)
and (f), are remarkably similar to the density profile obtained
from the ground state of a $4 \times 4$ cluster with open boundary
conditions in the $x$-direction and periodic boundary
conditions in the the $y$-direction,
the boundary conditions used by WS\cite{white98}.
The hole density as a function of the $x$-coordinate
for these three states is shown in Fig. \ref{fig:open}.
The states (e) and (f) are excited states of the periodic cluster \#2 in
Table \ref{table:clusters}.
The $4 \times 4$ cluster with open boundary conditions in the $x$-direction
can be generated from the periodic cluster \#2 by cutting all bonds along
one column.
Clearly, the excited states (e) and (f) essentially have an extra
node along one column.
The open boundary conditions in one direction causes the
ground state of this cluster to be very similar to excited states of
the periodic cluster.
The open boundary conditions select the striped state to be the ground state.

\begin{figure}[htb]
\epsfxsize=\figwidth\centerline{\epsffile{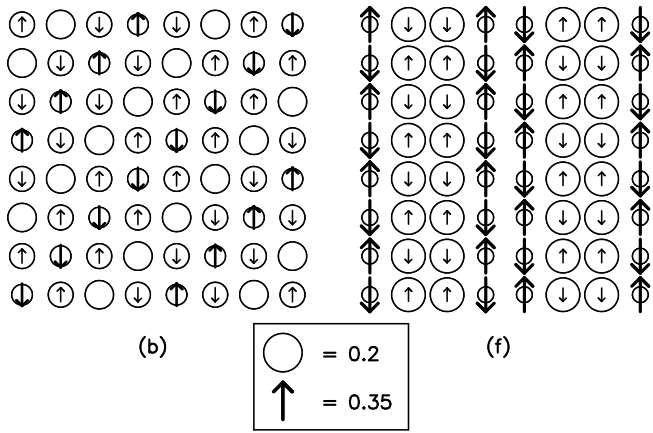}}
\vspace{.1in}
\caption{
Hole and spin structures of states (b) and (f) generated by applying
a small staggered magnetic field of $h=0.1$ to sites at the
boundary of the simulation cell to pin the spins.
The radius of each circle is proportional to the hole density
on the given site,
and the length of the arrows is proportional to $\langle S^z_i \rangle$.
These are {\em not} eigenstates
of the {\em t-J} Hamiltonian without the pinning fields,
but the spin correlations pictured are similar to the correlations
in the eigenstates (b) and (f).
State (e) has a spin and charge structure very similar to state (f)
and is not shown.
}
\label{fig:spin}
\end{figure}

None of the eigenstates has a spontaneous
spin density wave amplitude, but the spin correlations
are affected by the CDW.
One way to show the 
spin correlations is to apply small magnetic fields along the boundary of
the simulation cell,
as done by WS\cite{white98}.
This is shown for the diagonal and vertical stripe
states (b) and (f) in Fig.\ \ref{fig:spin}.
%In these states the spin order is in some way commensurate with the 
%CDW order.
%As would be expected,
The stripes are pinned so the sites
with the fields have maximum
electron density with the appropriate average polarization.
In both the diagonal and vertical cases,
the antiferromagnetic order in neighboring stripes is shifted
by $\pi$, as in the nickelates and cuprates\cite{nickelates,cuprates}.

%In addition, the stripe of Fig.\ \ref{fig:spin} are remarkably similar 
%to that obtained by WS on a larger-size system. Therefore, we conclude that 
%the method of WS did not converge to the true ground state of an infinite 
%size lattice but to a state which is a result of boundary conditions and
%finite-size effects. 

%Evidence of domain wall formation has been seen in hole-density
%correlation function studies of the {\em t-J} model\cite{prelovsek93}.
       
To conclude, we found stripes in the {\em t-J} model at a doping
of $x = \frac{1}{8}$, but only as 
excited states.
The ground state of the model for $J/t = 0.35$ is uniform.
The energy cost per site to form 
diagonal stripes similar to those found in the nickelates
is at least $\Delta_d \gtrsim 0.016t$, and
for vertical stripes similar to those in the cuprates the energy cost is
$\Delta_v \gtrsim 0.025t$.
% and has significantly different correlations than the striped phase.
At
%a hole density of $x = \frac{1}{8}$ and $J/t = 0.35$,
this doping and interaction strength
studies on larger systems have found that
the model is nearly phase separated\cite{hellberg97}.
The stripes seen experimentally could be the result of phase
separation frustrated by the Coulomb repulsion\cite{emery93}
and/or the coupling of the electrons to lattice distortions.
However, the stripe states are not ground states of the simple 
{\em t-J} model, as claimed in Ref.\ \cite{white98}. 

Our findings are in agreement with the recent work of Pryadko, Kivelson and 
Hone\cite{pryadko} who studied the interaction between localized holes in 
a weakly doped quantum antiferromagnet.  They find that stripes are 
unstable due to an attractive interaction between such domain walls. 

We thank R.E. Rudd and Steve Kivelson for numerous stimulating conversations.
This work was supported by the Office of Naval Research Grant
No.\ N00014-93-1-0189, and by the National Research Council. 
The calculations were performed on the SP2
at the DoD HPC Aeronautical Systems Center Major Shared Resource Center
at Wright Patterson Air Force Base.


\begin{references}

\bibitem[\dag]
{steve} Electronic address: hellberg@dave.nrl.navy.mil
\bibitem[\ddag]
{stratos} Electronic address: stratos@samos.martech.fsu.edu
\\Web address: www.stratos.fsu.edu

\bibitem{manousakis91}
E. Manousakis, Rev.\ Mod.\ Phys. {\bf 63},  1  (1991).

\bibitem{liu}
Z. Liu and E. Manousakis, Phys. Rev. {\bf B 45}, 2425 (1992).

\bibitem{wells}
B.O. Wells et al., Phys. Rev. Lett. {\bf 74}, 964 (1995).

\bibitem{boninsegni93}
M. Boninsegni and E. Manousakis, Phys.\ Rev.\ B {\bf 47},  11897  (1993).

\bibitem{emery90}
V.J. Emery, S.A. Kivelson, and H.Q. Lin, Phys.\ Rev.\ Lett. {\bf 64},  475
  (1990);
S.A. Kivelson, V.J. Emery, and H.Q. Lin, Phys.\ Rev.\ B
{\bf 42}, 6523 (1990);
S.A. Kivelson and V.J. Emery, 
in {\em Strongly correlated electronic materials:
the Los Alamos symposium, 1993},
edited by K.S. Bedell {\em et al.}\ (Addison-Wesley, Reading, CA, 1994).

\bibitem{hellberg97}
C.S. Hellberg and E. Manousakis, Phys.\ Rev.\ Lett. {\bf 78},  4609  (1997).

\bibitem{exactdiag}
E. Dagotto, Rev.\ Mod.\ Phys. {\bf 66},  763  (1994);
H. Fehske, V. Waas, H. R\"oder, and H. B\"uttner, Phys.\ Rev.\ B {\bf 44},
 8473  (1991);
D. Poilblanc, {\em ibid.} {\bf 52},  9201  (1995).

\bibitem{putikka}
M.U. Luchini,
{\em et al.},
% M. Ogata, W.O. Putikka, and T.M. Rice,
Physica C {\bf 185-189},  141  (1991);
W.O. Putikka, 
M.U. Luchini, and T.M. Rice, Phys.\ Rev.\ Lett. {\bf 68},  538  (1992).


\bibitem{kohno}
M. Kohno, Phys.\ Rev.\ B {\bf 55},  1435  (1997);

\bibitem{sorella}
M. Calandra, F. Becca and S. Sorella, e-print, Cond-mat/9810301.

\bibitem{siggia} B. Shraiman and E. Siggia, Phys. Rev. Lett. {\bf 60}, 740 (1988)
and {\it ibid.}  {\bf 61}, 467 (1988) 

\bibitem{boninsegni92}
M. Boninsegni and E. Manousakis, Phys.\ Rev.\ B {\bf 45}, 4877 (1992) and {\it ibid.},
 {\bf 46},  560  (1992).

\bibitem{emery93}
V.J. Emery and S. A. Kivelson, Physica {\bf C 209}, 597 (1993).

\bibitem{nickelates}
P. Wochner, J.M. Tranquada, D.J. Buttrey, and V. Sachan,
Phys.\ Rev.\ B {\bf 57}, 1066 (1998);
S.-H. Lee and S-W. Cheong, Phys.\ Rev.\ Lett. {\bf 79}, 2514 (1997);
J.M. Tranquada, D.J. Buttrey, V. Sachan, and J.E. Lorenzo,
{\it ibid} {\bf 73}, 1003 (1994);
V. Sachan,  D.J. Buttrey, J.M. Tranquada, J.E. Lorenzo, and G. Shirane,
Phys.\ Rev.\ B {\bf 51}, 12742 (1995);
J.M. Tranquada, J.E. Lorenzo, D.J. Buttrey, and V. Sachan,
{\it ibid} {\bf 52}, 3581 (1995).

\bibitem{cuprates}
M.v. Zimmermann, {\em et.\ al.}, Europhys.\ Lett., {\bf 41} 629 (1998);
J.M. Tranquada,  {\em et.\ al.}, Phys.\ Rev.\ Lett. {\bf 78}, 338 (1997);
J.M. Tranquada,  {\em et.\ al.}, Phys.\ Rev.\ B {\bf 54}, 7489 (1996);
J.M. Tranquada,  {\em et.\ al.}, Nature {\bf 375}, 561 (1995).

\bibitem{white98}
S.R. White and D.J. Scalapino, Phys.\ Rev.\ Lett. {\bf 80},  1272 (1998).

\bibitem{prelovsek93}
P. Prelov\v{s}ek and X. Zotos, Phys.\ Rev.\ B {\bf 47}, 5984 (1993).

\bibitem{zaanen94}
J. Zaanen and P.B. Littlewood, Phys.\ Rev.\ B {\bf 50}, 7222 (1994).

\bibitem{pryadko}
L. P. Pryadko, S. Kivelson and D. W. Hone, Phys. Rev. Lett, {\bf 80}, 5651 (1998).

\end{references}
\end{document}